\documentclass[aps,pra,twocolumn,superscriptaddress]{revtex4}
\usepackage{amsmath}
\usepackage{graphicx}
\usepackage[section]{placeins}
\usepackage{multirow} 
\usepackage{multirow}
\usepackage{subfigure}
\usepackage{color,soul}

\newcommand{\ket}[1]{|#1\rangle}

\newcommand{\eq}{\begin{equation}}
\newcommand{\fine}{\end{equation}}

\newcommand{\aac}{\'a}

\begin{document}
\title{Experimental generation and characterization of single-photon hybrid ququarts based on polarization-orbital angular momentum encoding}
\author{Eleonora Nagali}

\affiliation{Dipartimento di Fisica, Sapienza Universit\`{a} di Roma, Roma 00185, Italy}

\author{Linda Sansoni}

\affiliation{Dipartimento di Fisica, Sapienza Universit\`{a} di Roma, Roma 00185, Italy}

\author{Lorenzo Marrucci}

\affiliation{Dipartimento di Scienze Fisiche, Universit\`{a} di
Napoli ``Federico II'', Compl.\ Univ.\ di Monte S. Angelo, 80126
Napoli, Italy}

\affiliation{CNR-SPIN Coherentia, Compl.\ Univ.\ di Monte S. Angelo,
80126 Napoli, Italy}

\author{Enrico Santamato}

\affiliation{Dipartimento di Scienze Fisiche, Universit\`{a} di
Napoli ``Federico II'', Compl.\ Univ.\ di Monte S. Angelo, 80126
Napoli, Italy}

\affiliation{Consorzio Nazionale Interuniversitario per le Scienze
Fisiche della Materia, Napoli}

\author{Fabio Sciarrino}

\email{fabio.sciarrino@uniroma1.it}

\affiliation{Dipartimento di Fisica, Sapienza Universit\`{a} di Roma, Roma 00185, Italy}

\affiliation{Istituto Nazionale di Ottica Applicata, Firenze, Italy}

\begin{abstract}
High-dimensional quantum states, or \textit{qudits}, represent a promising resource in the quantum information field. Here we present the experimental generation of four-dimensional quantum states, or \textit{ququarts}, encoded in the polarization and orbital angular momentum of a single photon. Our novel technique, based on the q-plate device, allows to prepare and measure the ququart in all five mutually unbiased bases. We report the reconstruction of the four dimensional density matrix through the tomographic procedure for different ququart states.
\end{abstract}

\maketitle

\section{Introduction}
In quantum information theory the fundamental unit of information is a two-level system, the qubit. As in the classical information science, in
principle, all quantum information tasks can be performed based only
on qubits and on quantum gates working on qubits. For practical
reasons, however, there may be a significant advantage in
introducing the use of higher dimensional systems for encoding and
manipulating the quantum information. Such $d$-level quantum
systems, or \textit{qudits}, provide a natural extension of qubits
that has been shown to be suitable for prospective applications such
as quantum cryptography and computation \cite{Cerf02,Lany09}. The growing interest in qudit states lies on two main aspects. On one hand, the adoption of multi-dimensional states has been proven to increase the efficiency of
Bell-state measurements for quantum teleportation and to enhance the violation of Bell-type inequalities \cite{Coll02,Kasz00,Kasz02,Geno05,Vert09}. On the other hand, besides implications in fundamental quantum mechanics theory \cite{Wang03}, the qudit offers several advantages in the quantum information field \cite{Jozs00,DiVi03,Joo07}. Indeed, $d$-dimensional states are more robust against isotropic noise hence offering  higher transmission rates through communication channels \cite{Fuji03,Bish09}. Several quantum cryptographic protocols have been developed in order to capitalize the usefulness of these states, especially their capability to increase the security against eavesdropping attacks and the noise threshold that quantum key distribution protocols can tolerate \cite{Bech00,Bour01,Cerf02,Gisi02,Bruss02,Walb06,Kulik06,Kulik06b,Souz08,Shur09}. Furthermore, qudits offer advantages also for more efficient quantum gates \cite{Ralph07}, and for quantum information protocols \cite{Rana09,Bene09}. 

In the last years, many different implementations of qudits have
been proposed and demonstrated (see, e.g., \cite{Li08,Neel09}). In
quantum optics, implementations of \textit{qutrits} ($d=3$) and
\textit{ququarts} ($d=4$) states have been for example carried out
by exploiting two-photon polarization states
\cite{DAri05,Lany08,Baek08}. In such case, a natural basis for the
ququart Hilbert space of a photon pair is the following:
$\{\ket{H_{a},H_{b}},\ket{H_{a},V_{b}},\ket{V_{a},H_{b}},\ket{V_{a},V_{b}}\},
$
where $H$ and $V$ refer to the horizontal and vertical linear
polarization states, and $a,b$ denote the two photon modes. Hence a pair of photons can be exploited to encode a ququart state. When the two modes coincide ($a=b$), a qutrit system is obtained since the two states $\ket{H_{a},V_{b}}$ and $\ket{V_{a},H_{b}}$ become indistinguishable.
Different experiments on biphoton qudits have been performed by adopting the spontaneous parametric down-conversion process (SPDC): preparation and measurement of ququarts encoded in polarization and frequency degrees of freedom \cite{More06,Bogd06}, realization of polarization qutrit from nonmaximally entangled states \cite{Vall07}, entanglement between ququart systems \cite{Baek07}.
A biphoton implementation however has several limitations. First of all the rotated basis of the ququarts correspond to entangled states of the biphoton. Furthermore the implementation of unitary operation on a single quantum system requires an interaction between two photons, a task that can be achieved only probabilistically with linear optics. 
Finally the detection efficiency scales as $\eta^2$ where $\eta$ includes transmission losses and detection efficiency, thus low implementation rates are typically achieved.

An alternative approach to realize photon qudits is that based on
degrees of freedom other than polarization, within a single photon.
In this context, a particularly attractive choice is  the
orbital angular momentum (OAM) of light, providing a natural
discrete high-dimensional basis of photon quantum states within a
given longitudinal optical mode \cite{Fran08,Moli07,Naga09c}. Up to now, qudits with $d=3$ and $d=4$ encoded in photonic purely OAM systems have been generated \cite{Torr03,Vazi03} and employed in quantum communication \cite{Moli04}, quantum coin tossing \cite{Moli05}, quantum bit commitment \cite{Lang04}, and quantum key distribution \cite{Walb06}. However, despite its many potential advantages, the
use of OAM has been so far limited by technical difficulties arising
in the full manipulation and transmission of this degree of freedom.
For example, in contrast to polarization, OAM coherent
superpositions are affected by simple free-space propagation, owing
to the different Gouy-phases associated with different OAM values.\\

In the present manuscript we manipulate the polarization and OAM degrees of freedom to generate a ququart encoded in a single photon. Since we exploit two different degrees of freedom of the same particle, we refer to \textit{hybrid}-ququart states. Such results have been achieved through the q-plate device \cite{Marr06,Naga09}, which couples the spinorial (polarization) and orbital contributions of the angular momentum of photons. A complete characterization of the ququart states has been carried out by the quantum state tomography technique. 
\section{Ququart realization}
Let us consider the bidimensional OAM subspace with $m=\pm 2$, where $m$ denotes here
the OAM per photon along the beam axis in units of $\hbar$. We will denote such subspace as $o_2=\{\ket{+2},\ket{-2}\}$.
According to the nomenclature $\ket{\varphi,\phi}=\ket{\varphi}_{\pi}\ket{\phi}_{o_2}$, where $\ket{\cdot}_{\pi}$ and $\ket{\cdot}_{o_2}$ stand for the photon
quantum state `kets' in the polarization and OAM degrees of freedom, the logic ququart basis
\begin{equation*}
\{\ket{1},\ket{2},\ket{3},\ket{4}\}
\end{equation*}
can be re-written as:
\begin{equation*}
\{\ket{H,+2},\ket{H,-2},\ket{V,+2},\ket{V,-2}\}
\end{equation*}
where $H$ ($V$) refers to horizontal (vertical) polarization.
Following the same convention, the OAM equivalent of
the two basis linear polarizations $\ket{H}$ and $\ket{V}$ are then
defined as
\begin{eqnarray}
\ket{h}&=&\frac{1}{\sqrt{2}}(\ket{+2}+\ket{-2})\nonumber\\
\ket{v}&=&\frac{1}{i\sqrt{2}}(\ket{+2}-\ket{-2})
\end{eqnarray}
Finally, the $\pm45^{\circ}$ angle ``anti-diagonal'' and
``diagonal'' linear polarizations will be hereafter denoted with the
kets $\ket{A}=(\ket{H}+\ket{V})/\sqrt{2}$ and
$\ket{D}=(\ket{H}-\ket{V})/\sqrt{2}$, and the corresponding OAM
states are defined analogously:
\begin{eqnarray}
\ket{a}&=&\frac{1}{\sqrt{2}}(\ket{h}+\ket{v})=\frac{e^{-i\pi/4}}{\sqrt{2}}(\ket{+2}+i\ket{-2})\nonumber\\
\ket{d}&=&\frac{1}{\sqrt{2}}(\ket{h}-\ket{v})=\frac{e^{i\pi/4}}{\sqrt{2}}(\ket{+2}-i\ket{-2})
\end{eqnarray}
\begin{figure}[h]
\centering
\includegraphics[scale=.32]{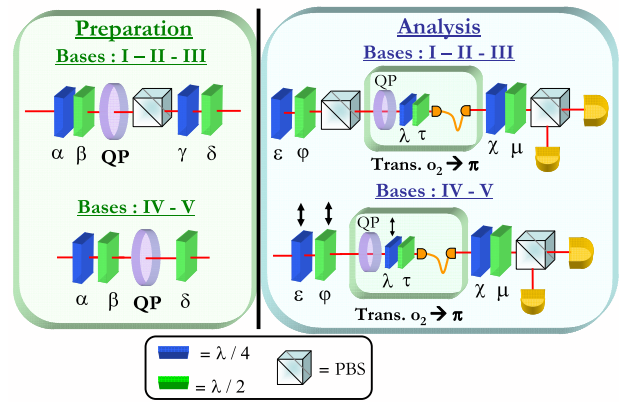}
\caption{(Color online) Schematic representation of the preparation and analysis setting stages of the ququart states adopted in our experiment. The preparation stages of the bases $I,II,III$ works with probability equal to $50\%$, however a deterministic implementation can be achieved by adopting the scheme proposed in \cite{Naga09b}. The preparation stages for states $IV,V$ are deterministic. All the analysis schemes are probabilistic with $p=50\%$, but the setup for $I,II,III$ can be again made deterministic by exploiting the $o_2\rightarrow\pi$ transferrer proposed in \cite{Naga09,Naga09b}.}
\label{setupdet}
\end{figure}
Since we deal with a four-dimensional Hilber space, the complete characterization of a ququart state is achieved by defining and measuring five mutually unbiased bases with four states each \cite{Klap03,Plan07}. Indeed a density matrix of a d-dimensional quantum system can be completely reconstructed from the measurements with respect to $d+1$ mutually unbiased bases \cite{Ivan81}.
The twenty states composing the five mutually unbiased bases, denoted as $I,II,III,IV,V$, are reported in Table I both in the ququart logic basis as well as in the OAM-polarization nomenclature. Let us stress that while the bases $I,II,III$ correspond to the measurement of $\sigma_z\otimes\widetilde{\sigma}_z,\sigma_x\otimes\widetilde{\sigma}_x,\sigma_y\otimes\widetilde{\sigma}_y$ on the two qubit systems -where the \textit{tilde} $\widetilde{}$ denotes the Pauli's operator for OAM states- the bases $IV,V$ refer to entangled states between polarization and OAM.
\section{Ququart manipulation via q-plate}
The main feature of the q-plate is its capability of coupling the spinorial (polarization) and orbital contributions of the angular momentum of photons. A q-plate (QP) is a birefringent slab having a suitably patterned transverse optical axis, with a topological singularity at its center \cite{Marr06}. 
\begin{table*}[t!!]
\begin{center}
{\footnotesize
\begin{tabular}{|c|c|c||c|c|c|c||c|c|c|c|c|c||c|}
\hline\hline
\multicolumn{3}{|c|}{Theory} & \multicolumn{11}{|c|}{Experimental implementation through the q-plate device}\\ \hline
\multicolumn{3}{|c|}{Ququart States} & \multicolumn{4}{|c|}{Preparation} & \multicolumn{6}{|c|}{Analysis} & $F_{exp}$\\ \hline
$$ & $Ququart \;\;Logic \;\; Bases$ & $OAM-\pi$ & $\alpha$ & $\beta$ & $\gamma$ & $\delta$ & $\epsilon$ & $\varphi$ & $\lambda$ & $\tau$ & $\chi$ & $\mu$ & $$\\ \hline \hline
\multirow{4}{*}{I} & $\ket{1}$ & $\ket{H,+2}$ & $-45$ & $0$ & $0$ & $0$ & $0$ & $0$ & $-45$ & $+45$ & $0$ & $0$ & $(99.9\pm0.4)\%$\\
 & $\ket{2}$ & $\ket{H,-2}$ & $+45$ & $0$ & $0$ & $0$ & $0$ & $0$ & $-45$ & $+45$ & $0$ & $+45$ & $(94.6\pm0.4)\%$\\
 & $\ket{3}$ & $\ket{V,+2}$ & $-45$ & $0$ & $0$ & $+45$ & $0$ & $+45$ & $-45$ & $+45$ & $0$ & $0$ & $(99.9\pm0.4)\%$\\
 & $\ket{4}$ & $\ket{V,-2}$ & $+45$ & $0$ & $0$ & $+45$ & $0$ & $+45$ & $-45$ & $+45$ & $0$ & $+45$ & $(95.8\pm0.4)\%$\\ \hline
\multirow{4}{*}{II} & $\frac{1}{2}(\ket{1}+\ket{2}+\ket{3}+\ket{4})$ & $\ket{A,h}$ & $0$ & $0$ & $0$ & $+22.5$ & $+45$ & $+22.5$ & $-45$ & $+45$ & $+45$ & $+22.5$ & $(95.0\pm0.4)\%$\\
 & $\frac{1}{2}(\ket{1}-\ket{2}+\ket{3}-\ket{4})$ & $\ket{A,v}$ & $0$ & $+45$ & $0$ & $+22.5$ & $+45$ & $+22.5$ & $-45$ & $+45$ & $+45$ & $+22.5$ & $(89.2\pm0.4)\%$\\
 & $\frac{1}{2}(\ket{1}+\ket{2}-\ket{3}-\ket{4})$ & $\ket{D,h}$ & $0$ & $0$ & $0$ & $-22.5$ & $+45$ & $-22.5$ & $-45$ & $+45$ & $+45$ & $-22.5$ & $(97.7\pm0.4)\%$\\
 & $\frac{1}{2}(\ket{1}-\ket{2}-\ket{3}+\ket{4})$ & $\ket{D,v}$ & $+45$ & $0$ & $0$ & $-22.5$ & $+45$ & $-22.5$ & $-45$ & $+45$ & $+45$ & $-22.5$ & $(95.0\pm0.4)\%$\\ \hline
\multirow{4}{*}{III} & $\frac{1}{2}(\ket{1}+i\ket{2}+i\ket{3}-\ket{4})$ & $\ket{R,a}$ & $0$ & $-22.5$ & $+45$ & $0$ & $+45$ & $0$ & $-45$ & $+45$ & $+45$ & $0$ & $(96.3\pm0.4)\%$\\
 & $\frac{1}{2}(\ket{1}-i\ket{2}+i\ket{3}+\ket{4})$ & $\ket{R,d}$ & $0$ & $+22.5$ & $+45$ & $0$ & $+45$ & $0$ & $-45$ & $+45$ & $+45$ & $0$ & $(95.7\pm0.4)\%$\\
 & $\frac{1}{2}(\ket{1}+i\ket{2}-i\ket{3}+\ket{4})$ & $\ket{L,a}$ & $0$ & $-22.5$ & $-45$ & $+45$ & $-45$ & $0$ & $-45$ & $+45$ & $-45$ & $0$ & $(94.1\pm0.4)\%$\\
 & $\frac{1}{2}(\ket{1}-i\ket{2}-i\ket{3}-\ket{4})$ & $\ket{L,d}$ & $0$ & $+22.5$ & $-45$ & $+45$ & $-45$ & $0$ & $-45$ & $+45$ & $-45$ & $0$ & $(94.5\pm0.4)\%$\\ \hline
\multirow{4}{*}{IV} & $\frac{1}{2}(\ket{1}+\ket{2}+i\ket{3}-i\ket{4})$ & $\frac{1}{\sqrt{2}}(\ket{R,+2}+\ket{L,-2})$ & $0$ & $0$ & $-$ & $-$ & $-$ & $-$ & $-$ & $0$ & $0$ & $0$ & $(84.8\pm0.4)\%$\\
 & $\frac{1}{2}(\ket{1}-\ket{2}+i\ket{3}+i\ket{4})$ & $\frac{1}{\sqrt{2}}(\ket{R,+2}-\ket{L,-2})$ & $0$ & $+45$ & $-$ & $-$ & $-$ & $-$ & $-$ & $0$ & $0$ & $+45$ & $(91.4\pm0.4)\%$\\
 & $\frac{1}{2}(\ket{1}+\ket{2}-i\ket{3}+i\ket{4})$ & $\frac{1}{\sqrt{2}}(\ket{L,+2}+\ket{R,-2})$ & $0$ & $0$ & $-$ & $+45$ & $-$ & $0$ & $-$ & $0$ & $0$ & $0$ & $(89.4\pm0.4)\%$\\
 & $\frac{1}{2}(\ket{1}-\ket{2}-i\ket{3}-i\ket{4})$ & $\frac{1}{\sqrt{2}}(\ket{L,+2}-\ket{R,-2})$ & $0$ & $+45$ & $-$ & $+45$ & $-$ & $0$ & $-$ & $0$ & $0$ & $+45$ & $(88.4\pm0.4)\%$\\ \hline
\multirow{4}{*}{V} & $\frac{1}{2}(\ket{1}+i\ket{2}+\ket{3}-i\ket{4})$ & $\frac{1}{\sqrt{2}}(\ket{H,a}+\ket{V,d})$ & $0$ & $+22.5$ & $-$ & $-$ & $+45$ & $-$ & $-45$ & $+45$ & $+45$ & $+22.5$ & $(89.7\pm0.4)\%$\\
 & $\frac{1}{2}(\ket{1}+i\ket{2}-\ket{3}+i\ket{4})$ & $\frac{1}{\sqrt{2}}(\ket{H,a}-\ket{V,d})$ & $0$ & $-22.5$ & $-$ & $-$ & $+45$ & $-$ & $-45$ & $+45$ & $+45$ & $-22.5$ & $(86.1\pm0.4)\%$\\
 & $\frac{1}{2}(\ket{1}-i\ket{2}+\ket{3}+i\ket{4})$ & $\frac{1}{\sqrt{2}}(\ket{H,d}+\ket{V,a})$ & $0$ & $+22.5$ & $-$ & $+45$ & $+45$ & $0$ & $-45$ & $+45$ & $+45$ & $+22.5$ & $(88.4\pm0.4)\%$\\
 & $\frac{1}{2}(\ket{1}-i\ket{2}-\ket{3}-i\ket{4})$ & $\frac{1}{\sqrt{2}}(\ket{H,d}-\ket{V,a})$ & $0$ & $-22.5$ & $-$ & $+45$ & $+45$ & $0$ & $-45$ & $+45$ & $+45$ & $-22.5$ & $(92.0\pm0.4)\%$\\ \hline\hline
\end{tabular}
}
\end{center}
\caption{Mutually unbiased bases for ququart states expressed in the logic ququart basis ${\ket{1},\ket{2},\ket{3},\ket{4}}$ and in term of polarization and orbital angular momentum. On the right side of the table are reported the different settings of the waveplates (see Fig.\ref{setupdet}) for the experimental preparation and analysis of the ququart states through the q-plate device. These settings correspond to the optical axis angle with respect to the horizontal laboratory axis, assuming that the input photon state is $\ket{H}_{\pi}\ket{0}_o$. On the last column on the right are reported the overall experimental fidelities including both the preparation and the measurement stages. The uncertainties have been evaluated according to the Poissonian
statistics of the photon counting.}
\end{table*}
The ``charge'' of this singularity is given by
an integer or half-integer number $q$, which is determined by the
(fixed) pattern of the optical axis. The birefringent retardation
$\delta_{QP}$ must instead be uniform across the device. 
For $\delta_{QP}=\pi$, a QP modifies the OAM state $m$ of a light beam
crossing it, imposing a variation $\Delta{m}={\pm}2q$ whose sign
depends on the input polarization, positive for left-circular and
negative for right-circular. On the polarization degree of freedom, the q-plate acts as a half-waveplate. 
In the present work, we use QPs with charge
$q=1$ and retardation $\delta=\pi$. Hence, an input TEM$_{00}$ mode (having
$m=0$) is converted into a beam with $m=\pm2$:
\begin{eqnarray}
\ket{L}_{\pi}\ket{0}_{o} & \xrightarrow{QP} & \ket{R}_{\pi}\ket{+2}_{o_2} \nonumber\\
\ket{R}_{\pi}\ket{0}_{o} & \xrightarrow{QP} &
\ket{L}_{\pi}\ket{-2}_{o_2} \label{eqqplate}
\end{eqnarray}
where $L$ and $R$ denote the left and right circular polarization
states, respectively. It has been experimentally demonstrated that any coherent superposition of the two input states given in Eq.(\ref{eqqplate}) is preserved by the QP
transformation, leading to the equivalent superposition of the
corresponding output states \cite{Naga09}.
Hence the QP can be easily employed to generate single-particle entanglement of $\pi$ and OAM degrees of freedom, a property that can be exploited to generate ququart states in the entangled mutually unbiased bases $IV$ and $V$. Such property, experimentally verified in \cite{Naga09}, can be synthetically expressed as follows:
\begin{equation}
\left.
\begin{array}{c}
\left| H\right\rangle _{\pi }\left| 0\right\rangle _{o} \\
\left| V\right\rangle _{\pi }\left| 0\right\rangle _{o}
\end{array}
\right\} \overset{QP}{\leftrightarrow}\frac{1}{\sqrt{2}}(\ket{L}_{\pi}\ket{-2}_{o_2}\pm\ket{R}_{\pi}\ket{+2}_{o_2})
\label{entangled}
\end{equation}
This is an entangled state between two qubits encoded in different degrees of freedom belonging to the mutually unbiased basis $IV$ for ququart states. Thus the ability of the q-plate to \textit{entangle}-\textit{disentangle} the OAM-polarization degrees of freedom will be exploited for the preparation as well as for the measurement of the ququart states.
For the preparation and analysis stages, let us refer to Fig.\ref{setupdet}.\\
\textbf{Preparation stage.} 
The states belonging to bases $I,II,III$ are generated through the transferrer $\pi\rightarrow o_2$, which achieves the
transformation $\ket{\varphi}_{\pi}\ket{0}_{o}\rightarrow\ket{H}_{\pi}\ket{\varphi}_{o_2} $, where $\ket{0}_o$ denotes the zero-OAM state. This transferrer, presented in \cite{Naga09,Naga09b}, is based on two waveplates, a q-plate and a polarizing beam-splitter, thus allowing the coherent transfer of the information initially encoded in the polarization to the OAM degree of freedom. After the transferrer $\pi\rightarrow o_2$ it is possible to encode new information in the polarization by adopting two waveplates. The preparation stage for states belonging to bases $IV,V$ is based on two waveplates and a q-plate, which allow to achieve transformations analogous to the one in Eq.(\ref{entangled}), depending on the polarization state initially encoded through the waveplates. The half-waveplate $\delta$ (shown in Fig.\ref{setupdet}) can be inserted depending on the specific state to be prepared.
In Table I are reported the rotations of the waveplates in the preparation stage to be applied for the generation of all the ququart states. We note that by this method we can generate
all ququarts belonging to the five unbiased bases, but not an
arbitrary ququart. The latter could be obtained by implementing the
more complex universal unitary gate proposed in Ref.\ \cite{Slus09}.
However, to demonstrate the generality of any quantum protocol it is
enough to test it on all states belonging to a complete set of
unbiased bases, as those generated in the present work.

\textbf{Analysis stage}
The measurement procedure of the ququart states depends on the bases to be analysed. The schemes are reported in Fig.\ref{setupdet}, right side.
The states belonging to bases $I,II,III$ have been measured first in the polarization contribution of the wavefunction through a standard polarization analysis set, composed of two waveplates and a polarizing beam splitter (PBS). Then the orbital contribution has been analysed adopting the quantum transferrer $o_2\rightarrow\pi$, which achieves the
transformation $\ket{H}_{\pi}\ket{\varphi}_{o_2}\Rightarrow
\ket{\varphi}_{\pi}\ket{0}_{o}$ \cite{Naga09,Naga09b}. Such transferrer is based on a q-plate, two waveplates, and a single-mode fiber which selects only the orbital contributions with $m=0$. Hence this analysis stage is a probabilistic one, with $p=0.5$. After the transferrer the state $\ket{\varphi}_{\pi}\ket{0}_{o}$ has been measured through a standard polarization analysis set. Two multi-mode fibers collect the output signals of the PBS and then send them to the detectors, as described in detail in the experimental section.
States belonging to bases $IV,V$ have been mesured exploiting the q-plate capability of disentagling the polarization from the OAM degree of freedom. Thus, depending on the state to be analysed, two waveplates have been inserted in order to manipulate the polarization component of the wavefunction, and then the state has been sent through the $o_2\rightarrow\pi$ transferrer. Finally the output state from the single mode fiber has been measured through a standard polarization analysis set, as for states belonging to bases $I,II,III$. Let us stress that also this detection stage is a probabilistic one, with $p=0.5$.

Nevertheless it is possible to realize two fully
\emph{deterministic} transferrers $\pi\rightarrow o_2$ and $o_2\rightarrow\pi$ at the price of a more complex optical layout, based on a q-plate and a
Sagnac interferometer, as shown in \cite{Naga09b}. This result can be applied in order to achieve a deterministic preparation stage for all the bases $I,II,III,IV,V$ and a deterministic measurement stage for bases $I,II,III$.
\begin{figure}[h]
\centering
\includegraphics[scale=.28]{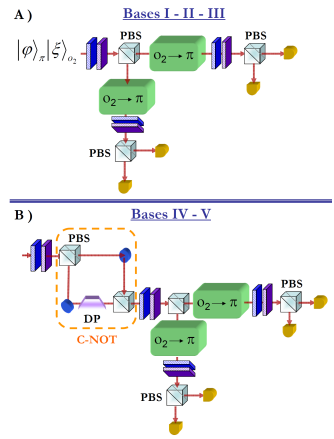}
\caption{(Color online) Schematic representation of the deterministic analysis setting stages of the ququart states belonging to \textbf{(A)} bases $I,II,III$ and \textbf{(B)} to bases $IV,V$. The $o_2\rightarrow\pi$ transferrer in both schemes works deterministically, as proposed in \cite{Naga09b}.}
\label{misdet}
\end{figure}
In Fig.\ref{misdet} we propose two schemes in order to analyse deterministically all the ququart states belonging to the five mutually unbiased bases. In particular states of bases $IV,V$ can be measured with $p=1$ by inserting a $C-NOT$ scheme where the polarization controls the OAM degree of freedom. Depending on the $\pi$ contribution of the wavefuntion, in the Sagnac interferometer a $\sigma_z$ operator, implemented through a Dove's prism (DP) properly rotated, acts on the orbital degree of freedom. Hence the state is analysed in polarization and then in OAM by the deterministic $o_2\rightarrow\pi$ trasnferrer. It is worth noting that the schemes here proposed can be adopted in order to carry out a quantum Bell's state measurement of an hybrid ququart state encoded in polarization and OAM.
\section{Experimental characterization}
\begin{figure}[h]
\begin{center}
\includegraphics[width=8cm]{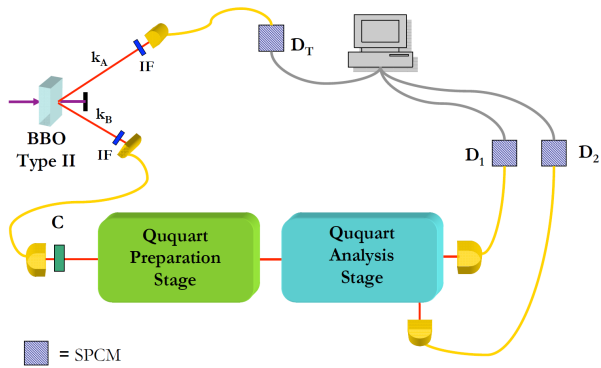}
\caption{(Color online) Experimental setup adopted for the generation and characterization of ququart states encoded in the polarization and orbital angular momentum of a single photon.}
\label{setup}
\end{center}
\end{figure}

The experimental layout is shown in Fig.\ref{setup}. A $\beta$-barium borate
crystal (BBO) cut for type-II phase matching, pumped by the
second harmonic of a Ti:Sa mode-locked laser beam, generates via
spontaneous parametric fluorescence photon pairs on modes $k_A$ and $k_B$ with linear polarization, wavelength $\lambda=795$ nm, and pulse bandwidth $\Delta\lambda=4.5$ nm, as determined by two interference filters (IF). The coincidence rate of the source is equal to 8 kHz.
The photon generated on mode $k_A$ is detected at $D_T$, thus acting as a trigger on the single-photon generation. The photon generated on mode $k_B$ is
delivered to the setup via a single mode fiber, thus defining its transverse spatial mode to a pure TEM$_{00}$, corresponding to OAM $m=0$. After the fiber output, two waveplates compensate (C) the polarization rotation introduced by the fiber. Then the ququart is encoded in the single photon polarization and OAM through the ququart preparation stage, whose structure depends on the state to be generated, as described in the previous section.
After the ququart analysis stage, the output photons have been coupled to  single mode fibers and then detected by two single-photon counter modules (SPCM) connected to the coincidence box (CB), which records the coincidence counts between $[D_1,D_T]$ and $[D_2,D_T]$.
For the detailed representation of the preparation and analysis ququart stages, let us refer to Fig.\ref{setupdet}. 

As first step, we have estimated the fidelities for all quantum states of the 5
mutually unbiased basis. The experimental fidelities $F_{exp}$ are reported in Table I. For every input state $\ket{\varphi}$, $F_{exp}$ has been estimated as $F_{exp}=\frac{p(\ket{\varphi})}{\sum_{i}p(\ket{\psi_i})}$ where $p(\ket{\psi_i})$ is
the probability to detect $\ket{\psi_i}$. The average value $\bar{F}$ for every basis reads:

\begin{equation*}
\begin{tabular}{|l|l|}
\hline
Basis & \;\;\;\;$\bar{F}(\%)$ \\ \hline
I & $97.6\pm0.4$ \\ \hline
II & $94.2\pm0.4$ \\ \hline
III & $95.2\pm0.4$ \\ \hline
IV & $88.5\pm0.4$ \\ \hline
V & $89.1\pm0.4$ \\ \hline
\end{tabular}
\end{equation*}

These values take into account the fidelity of both the preparation and measurement steps.
Since these two processes are fairly symmetrical, the fidelities associated to the single preparation/measurement stage can be estimated as the square root of the previous values. 

As can be observed from the previous table, there is a slight difference ($\sim 6\%$) between the mean fidelity value of states belonging to the first three basis, and the one related to states of the $IV$ and $V$ basis. Indeed a discrepancy of around $3\%$ lies on the present limitation of the efficiency, which implies small random contributions to the polarization after the generation of the ququart states. Since for the analysis of entangled states between $\pi$-OAM the polarizing beam splitter was removed, such unconverted contributions are not filtered and thus affect the final fidelity. The remaining $3\%$ is due to phase effects between different contributions in the entangled $\pi$-OAM wavefunction, as well as a slight misalignment respect to the separable states-experimental setup.
\section{Quantum state tomography of ququart}
The ability to prepare the ququart codified in the OAM-polarization space
in all the mutually unbiased bases has been experimentally verified by reconstructing their density matrices through quantum state tomography.

The experimental density matrix associated to a ququart can be expressed as 
\begin{equation}
\widehat{\rho }=\frac{1}{4}\sum_{j=1}^{d^{2}=16}r_{j}\widehat{\lambda }_{j}
\end{equation}
where $\{\widehat{\lambda }_{j}\}$ is a complete operatorial set of hermitian
generators and $r_{j}=\left\langle \widehat{\lambda }_{j}\right\rangle =Tr[%
\widehat{\rho }\widehat{\lambda }_{j}]$.

\begin{figure}[t!!]
\begin{center}
\includegraphics[scale=.3]{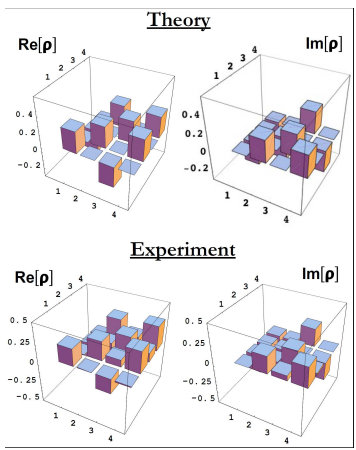}
\caption{(Color online) Theoretical and experimental density matrix of a ququart state belonging to the fifth mutually unbiased basis. The state reported is $(2^{1/2})(\ket{H,d}-\ket{V,a})$. On axis are reported the states $\ket{1},\ket{2},\ket{3},\ket{4}$, schematically written as $1,2,3,4$. On the ordinate are reported the elements of the density matrix.}
\label{tomoent}
\end{center}
\end{figure}

The set of operators $\{\widehat{\lambda }_{j}\}$ for the ququart here
considered can be obtained starting from the generators of $SU(2)$ as $%
SU(2)\otimes SU(2)$: $\widehat{\sigma }_{i}^{\pi}\otimes \widehat{\sigma }_{j}^{o_2}.$
The tomography reconstruction obtained in the following way requires the estimation
of 16 operators \cite{Jame01} through 36 separable measurements on the polarization-OAM subspaces. 
\begin{figure}[h]
\begin{center}
\includegraphics[scale=.25]{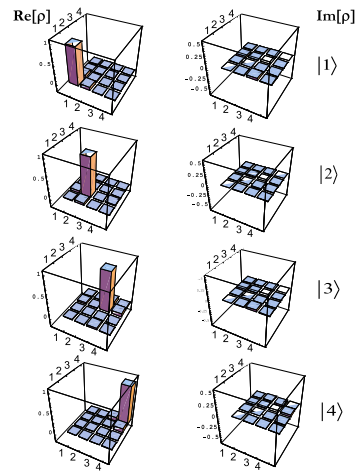}
\caption{(Color online) Experimental density matrices of ququart states belonging to the first mutually unbiased basis. On axis are reported the states $\ket{1},\ket{2},\ket{3},\ket{4}$, schematically written as $1,2,3,4$. On the ordinate are reported the elements of the density matrix.}
\label{tomodis}
\end{center}
\end{figure}
\begin{equation*}
\begin{tabular}{|c|c|}
\hline
$\ \ \ \ \widehat{\lambda }_{1}=\widehat{\sigma }_{X}\otimes \widehat{I}$ \
\ \ \  &  $\ \ \ \ \widehat{\lambda }_{9}=\widehat{\sigma }_{X}\otimes 
\widehat{\sigma }_{Y}$ \ \ \ \  \\ \hline
$\widehat{\lambda }_{2}=\widehat{\sigma }_{Y}\otimes \widehat{I}$ & $%
\widehat{\lambda }_{10}=\widehat{\sigma }_{Y}\otimes \widehat{\sigma }_{Y}$
\\ \hline
$\widehat{\lambda }_{3}=\widehat{\sigma }_{Z}\otimes \widehat{I}$ & $%
\widehat{\lambda }_{11}=\widehat{\sigma }_{Z}\otimes \widehat{\sigma }_{Y}$
\\ \hline
$\widehat{\lambda }_{4}=\widehat{I}\otimes \widehat{\sigma }_{X}$ & $%
\widehat{\lambda }_{12}=\widehat{I}\otimes \widehat{\sigma }_{Z}$ \\ \hline
$\widehat{\lambda }_{5}=\widehat{\sigma }_{X}\otimes \widehat{\sigma }_{X}$
& $\widehat{\lambda }_{13}=\widehat{\sigma }_{X}\otimes \widehat{\sigma }_{Z}
$ \\ \hline
$\widehat{\lambda }_{6}=\widehat{\sigma }_{Y}\otimes \widehat{\sigma }_{X}$
& $\widehat{\lambda }_{14}=\widehat{\sigma }_{Y}\otimes \widehat{\sigma }_{Z}
$ \\ \hline
$\widehat{\lambda }_{7}=\widehat{\sigma }_{Z}\otimes \widehat{\sigma }_{X}$
& $\widehat{\lambda }_{15}=\widehat{\sigma }_{Z}\otimes \widehat{\sigma }_{Z}
$ \\ \hline
$\widehat{\lambda }_{8}=\widehat{I}\otimes \widehat{\sigma }_{Y}$ & $%
\widehat{\lambda }_{16}=\widehat{I}\otimes \widehat{I}$ \\ \hline
\end{tabular}
\end{equation*}
We carried out the reconstruction of $\rho$ for different states belonging to all the 
basis previously introduced. Examples of the experimental results are reported in Fig.\ref{tomodis}-\ref{tomoent}. The accordance between theory and experiment can be appreciated by looking at the tomographies as well as through the fidelity $F=\left\langle \phi \right| \rho \left| \phi
\right\rangle $ and the linear entropy $S_L$ values, theoretically equal to zero, with the ideal states evaluated for some states belonging to the five mutually unbiased basis:\\ 
\begin{center}
\begin{tabular}{|l|l|l|}
\hline
\;\;\;\;\;\;\;Input states & $\;\;\;F(\%)$ & $\;\;\;\;\;\;\;\;S_{L}$ \\ \hline
$\;\;\;\;\;\;\;\;\;\;\;\ket{H,-2}$ & $98.5\pm0.3$ & $0.058\pm 0.002$ \\ \hline
$\;\;\;\;\;\;\;\;\;\;\;\;\ket{A,v}$ & $90.5\pm0.3$ &  $0.046\pm 0.002$ \\ \hline
$\;\;\;\;\;\;\;\;\;\;\;\;\ket{L,a}$ & $95.4\pm0.3$ & $0.097\pm 0.002$ \\ \hline
$\frac{1}{\sqrt{2}}(\ket{R,+2}-\ket{L,-2})$ & $93.0\pm0.3$ &  $0.092\pm 0.002$ \\ \hline
$\frac{1}{\sqrt{2}}(\ket{H,a}-\ket{V,d})$ & $92.0\pm0.3$ & $0.094\pm 0.002$ \\ \hline
\end{tabular}
\end{center}


Let us stress that the adoption of the q-plate has allowed the reconstruction of the density matrix also for the states belonging to the fourth and fifth basis, related to entangled states between orbital angular momentum and polarization of a single photon.
The previous approach requires the measurements of $16$ operators and the
corresponding $64$ eigenstates. A more efficient quantum state
reconstruction could be achieved by adopting all the mutually unbiased bases as just proposed and demonstrated with a 2-photon system by \cite{Adam08}. The reduced number of bases in a mutually unbiased basis tomography allows to reduce by a factor $5/9$ the number of measurements to achieve the same reconstruction accuracy. Moreover, this technique turned out to be even more efficient for entangled states since it directly estimates the strength of correlations in entangled bases ($IV$ and $V$) \cite{Adam08}.

\section{Conclusion}
In this paper we have experimentally carried out the encoding of a ququart state in the polarization and orbital angular momentum of a single photon. The states generated have been fully characterized through the quantum state tomography by measuring all the five mutually unbiased bases expected in a four-dimensional Hilbert space. The capacity to manipulate with high reliability all quqart states encoded in a single photon enforce the achievement of the quantum state engineering of ququart to direct towards the implementation of new quantum information protocols and more robust communication procedures. In particular, single-photon encoded ququart states could be adopted for the implementation of a quantum key distribution protocol without a reference frame recentely proposed \cite{Souz08,Shur09}, as well as for the implementation of a universal unitary gate \cite{Slus09}. Finally the advantage of encoding a ququart state in a single photon could be exploited for quantum information protocols like the optimal quantum cloning of quantum states with dimension $d>2$ \cite{Naga09c}.

E. N. and F. S. acknowledge support by FARI project and Finanziamento Ateneo 2009 of Sapienza Universit{\aac} di Roma.

\end{document}